\documentclass[nofootinbib,prd]{revtex4}%
\usepackage{amsmath}
\usepackage{amsfonts}
\usepackage{amssymb}
\usepackage{graphicx}
\usepackage{mathrsfs}
\usepackage{yfonts}
\usepackage{tipa}
\usepackage{bm}
\usepackage{bbm}
\usepackage{hyperref}
\usepackage{latexsym}
\usepackage{xcolor}
\usepackage[iso-8859-7]{inputenc}%
\begin{document}
\title{Rotating Casimir Wormholes}
\author{Remo Garattini}
\email{remo.garattini@unibg.it}
\affiliation{Universit\`{a} degli Studi di Bergamo, Dipartimento di Ingegneria e Scienze
Applicate, Viale Marconi 5, 24044 Dalmine (Bergamo) Italy }
\affiliation{I.N.F.N. - sezione di Milano, Milan, Italy.}
\author{Athanasios G. Tzikas}
\email{athanasios.tzikas@unibg.it}
\affiliation{Universit\`{a} degli Studi di Bergamo, Dipartimento di Ingegneria e Scienze
Applicate,Viale Marconi 5, 24044 Dalmine (Bergamo) Italy. }

\begin{abstract}
A Casimir Wormhole is a Traversable Wormhole powered by a Casimir energy
source within a static reference frame. A natural extension of this system is
the inclusion of rotation. We will explore two basic configurations: one with
radially varying Casimir plates and another with parametrically fixed plates.
In both cases, we will show that rotations do not alter the structure of a
Casimir wormhole, and the behavior observed in a static frame is reaffirmed.
Since the case with radially varying plates predicts a constant angular
velocity as a solution, we must introduce an exponential cut-off and an
additional scale to prevent rotations at infinity. This adjustment is not
necessary when the plates are kept parametrically fixed. Moreover, the
consistency of the Einstein Field Equations is ensured with the help of an
additional source without an accompanying energy density, which we interpret
as a thermal stress tensor.

\end{abstract}
\maketitle

\section{Introduction}

A Casimir Wormhole (CW) \cite{CW} is a solution of the Einstein Field
Equations (EFE) representing a Traversable Wormhole (TW) with a source
described by a Stress-Energy Tensor (SET) of the form
\begin{equation}
T_{\nu}^{\mu}|_{\mathrm{C}}=-\frac{\hbar c\pi^{2}}{720d^{4}}\mathrm{diag}%
\left(  -1,3,-1,-1\right)  .\label{Casimir}%
\end{equation}
The distance $d$ represents the separation between two parallel, closely
spaced, uncharged metallic plates in vacuum at almost zero temperature. To
obtain a CW, one has to replace the parametrically fixed distance $d$ with a
radially varying variable $r$. In other words, the quadruple $\left(
\rho(r),p_{r}(r),p_{t}(r),p_{t}(r)\right)  $, which includes the energy
density $\rho(r)$, the radial pressure $p_{r}(r)$ and the tangential pressure
$p_{t}(r)$, represents the physical source as follows:
\begin{equation}
T_{\nu}^{\mu}|_{\mathrm{C}}=-\frac{r_{1}^{2}}{\kappa r^{4}}\mathrm{diag}%
\left(  -1,3,-1,-1\right)  \qquad\mathrm{with}\qquad r_{1}^{2}=\frac{\pi^{3}%
}{90}\left(  \frac{\hbar G}{c^{3}}\right)  =\frac{\pi^{3}}{90}\ell
_{\mathrm{P}}^{2}\,.\label{CSET}%
\end{equation}
By using the conventional line element of a TW \cite{MT,MTY,Visser}
\begin{equation}
\mathrm{d}s^{2}=-e^{2\Phi\left(  r\right)  }\,\mathrm{d}t^{2}+\frac
{\mathrm{d}r^{2}}{1-b(r)/r}+r^{2}\,(\mathrm{d}\theta^{2}+\sin^{2}{\theta
}\,\mathrm{d}\varphi^{2})\,,\label{MT_metric}%
\end{equation}
the EFE can be written as\footnote{Throughout the paper, the field equations
are not examined in an orthonormal frame. Instead, Schwarzschild-like
coordinates ($t,r,\theta,\varphi$) are used for convenience in solving
equations.}\footnote{We must observe that by assuming the form of the SET \eqref{Casimir}, a consistent calculation is obtained if we adopt the configuration considered in  \cite{MTY}, where the plates have a spherical form. As noted in \cite{MTY}, this approximation introduces an error that remains small when we are very close to the throat, which is precisely the condition required in this paper. In this way, even with the inclusion of rotation, the physical setup remains essentially the same.}
\begin{equation}
\frac{b^{\prime}\left(  r\right)  }{r^{2}}=\kappa\rho\left(  r\right)
\label{rho}%
\end{equation}%
\begin{equation}
\frac{2}{r}\left(  1-\frac{b\left(  r\right)  }{r}\right)  \Phi^{\prime
}\left(  r\right)  -\frac{b\left(  r\right)  }{r^{3}}=\kappa p_{r}\left(
r\right)  \label{pr}%
\end{equation}%
\begin{equation}
\left(  1-\frac{b\left(  r\right)  }{r}\right)  \left[  \Phi^{\prime\prime
}\left(  r\right)  +\Phi^{\prime}\left(  r\right)  \left(  \Phi^{\prime
}\left(  r\right)  +\frac{1}{r}\right)  \right]  -\frac{b^{\prime}\left(
r\right)  r-b\left(  r\right)  }{2r^{2}}\left(  \Phi^{\prime}\left(  r\right)
+\frac{1}{r}\right)  =\kappa p_{t}(r)\,\label{pt}%
\end{equation}
in which $b(r)$ and $\Phi(r)$ denote the shape function and the redshift
function, respectively. The EFE can be supplemented with the expression for
the conservation of the SET, which is written in the same reference frame as
\begin{equation}
p_{r}^{\prime}\left(  r\right)  =\frac{2}{r}\left(  p_{t}\left(  r\right)
-p_{r}\left(  r\right)  \right)  -\left(  \rho\left(  r\right)  +p_{r}\left(
r\right)  \right)  \Phi^{\prime}\left(  r\right)  .
\end{equation}
The solution to the previous field equations with the Casimir source
\eqref{CSET} is given by
\begin{align}
\Phi\left(  r\right)   &  =\ln\left(  \frac{3r}{3r+r_{0}}\right)
\label{PhiCW}\\
b\left(  r\right)   &  =\frac{2}{3}r_{0}+\frac{r_{0}^{2}}{3r}\,.\label{b(r)CW}%
\end{align}
Such a solution possesses the property of satisfying the following
relationship between the energy density and the radial pressure
\begin{equation}
p_{r}(r)=\omega\rho\left(  r\right)  \quad(\omega=3)
\end{equation}
with
\begin{equation}
r_{0}=\sqrt{3}\,r_{1}\,.\label{r0}%
\end{equation}
It is interesting to observe that a solution can also be derived from the
original form of the SET represented by \eqref{Casimir}. To this purpose, we
consider the following setting%
\begin{align}
\Phi\left(  r\right)   &  =0\,\label{RFfp}\\
b\left(  r\right)   &  =r_{0}-\frac{r_{1}^{2}}{3d^{4}}\left(  r^{3}-r_{0}%
^{3}\right)  .\label{b(r)fp}%
\end{align}
As discussed also in Refs.\cite{CW,GABTW}, the shape function \eqref{b(r)fp}
does not represent a TW in the strict sense. However, it can be observed that
there exists a value $\bar{r}$ such that%
\begin{equation}
b\left(  \bar{r}\right)  =0\qquad\Longleftrightarrow\qquad\bar{r}%
=r_{0}\sqrt[3]{1+\frac{3d^{4}}{r_{0}^{2}r_{1}^{2}}}\,.\label{rb}%
\end{equation}
Therefore, a TW can be defined for $\Phi(r)=0$ in the interval $\left[
r_{0},+\infty\right)  $ and for
\[
b(r)=\left\{
\begin{array}
[c]{cc}%
r_{0}-\frac{r_{1}^{2}}{3d^{4}}\left(  r^{3}-r_{0}^{3}\right)  \quad &
r_{0}\,\leq r<\bar{r}\\
& \\
0 & r\geq\bar{r}\,.
\end{array}\label{CWd}
\right.
\]
The previous profile can be justified by imposing an equation of state (EoS)
of the form%
\[
\omega(r)=\left\{
\begin{array}
[c]{cc}%
\frac{d^{4}}{r_{1}^{2}r^{3}}\left(  r_{0}-\frac{r_{1}^{2}}{3d^{4}}\left(
r^{3}-r_{0}^{3}\right)  \right)  \quad & r_{0}\,\leq r<\bar{r}\\
& \\
0 & r\geq\bar{r}\,
\end{array}
\right.
\]
with $\omega(r_{0})=\frac{d^{4}}{r_{1}^{2}r_{0}^{2}}\,$. If we compare the
above expression for $\omega(r)$ with%
\begin{equation}
\frac{p_{r}\left(  r\right)  }{\rho\left(  r\right)  }=3=\omega(r)\,,
\end{equation}
we observe that this condition is satisfied only at the throat, where its
value is
\begin{equation}
r_{0}=\frac{\sqrt{3}d^{2}}{3r_{1}}\,.\label{r0d}%
\end{equation}
Note that, this time, Eq.~\eqref{r0d} predicts an enormously large throat
size. It is interesting to observe that the solution in \eqref{r0d} allows for
the evaluation of the location of $\bar{r}$. Indeed, substituting \eqref{r0d}
into \eqref{rb}, one obtains
\begin{equation}
\bar{r}=\sqrt[3]{10}\,r_{0}\,.\label{rbar}%
\end{equation}
This implies that the boundary $\bar{r}$ is located very close to the throat.
Note also that the plates are inside this boundary as shown in
Eq.$\left(  \ref{rb}\right)  $. One can observe that the structure of the TW
defined by \eqref{CWd} resembles an Absurdly Benign Traversable Wormhole
(ABTW) \cite{GABTW}, with the distinction that the energy density is
non-vanishing outside the boundary $\bar{r}$. We remind the reader that an
ABTW is defined by
\begin{equation}
b(r)=\left\{
\begin{array}
[c]{cc}%
r_{0}\left(  1-\lambda\left(  r-r_{0}\right)  \right)  ^{2}\qquad & r_{0}\leq
r<\bar{r}\\
& \\
0 & r\geq\bar{r}\,
\end{array}
\right.
\end{equation}
with $\Phi(r)=0$ everywhere and $\bar{r}=r_{0}+1/\lambda\,$. This means that
we have to assume that the energy density must be zero for $r>\bar{r}\,$. In
other words, outside $\bar{r}$, we recover Minkowski spacetime. Such an
assumption is not unusual, as the negativity of the energy density appears
only between the plates and not outside of them. A comment on this point is in order: in his book, Visser \cite{Visser} proposed a realistic model for
the total Stress-Energy Tensor (SET), represented by%
\begin{gather}
T_{\sigma}^{\mu\nu}=\sigma\hat{t}^{\mu}\hat{t}^{\nu}\left[  \delta\left(
z\right)  +\delta\left(  z-a\right)  \right] 
+\Theta\left(  z\right)  \Theta\left(  a-z\right)  \frac{\hbar c\pi^{2}%
}{720a^{4}}\left[  \eta^{\mu\nu}-4\hat{z}^{\mu}\hat{z}^{\nu}\right]  \,
\label{TCas}%
\end{gather}
where $\hat{t}^{\mu}$ is a unit time-like vector, $\hat{z}^{\mu}$ is a normal
vector to the plates and $\sigma$ is the mass density of the plates. He concluded that the mass of the plates compensates for the negative energy density to such an extent that it prevents the creation of a TW.  However, this does not imply that the general structure of a TW cannot be investigated. Rather, one could explore the conditions necessary to minimize the positive effects  of the mass of the plates. A more suitable approach for addressing this issue could be a thin-shell model, which is beyond the scope of this paper. 

What we have introduced is
valid for a static TW, but what happens when we introduce rotations?
Historically, the first proposal of a rotating TW was made by Teo \cite{Teo},
while an interesting analysis of the properties of the SET describing such a
wormhole was performed by Perez Bergliaffa and Hibberd \cite{PBH}. Kuhfittig
found a solution that exhibits a less severe violation of the Weak Energy
Condition (WEC) compared to the static case \cite{Kuhfittig}. Kim considered
scalar perturbations around a rotating TW, while an investigation of slow
rotation was conducted by Kashargin and Sushkov \cite{KS,KS1}. In this paper,
we aim to answer the following question: \textit{Is it possible to introduce
rotations in such a way that a Casimir wormhole is obtained in the
non-rotating limit, even in the case of parametrically fixed plates?} To
answer this question, we need to construct a spacetime that describes such a
rotating wormhole. The rest of the paper is organized as follows: in
Sec.~\ref{p2}, we examine the structure of both the rotating metric and the
SET. In Sec.~\ref{p3}, we examine the EFE for a rotating TW with the Casimir
plates positioned at a distance either parametrically fixed or radially
varying. We summarize and conclude in Sec.~\ref{p4}. Units in which $\hbar=c=k=1$ are used throughout the paper
and will be reintroduced whenever it is necessary.

\section{Setting up the Rotating Space-Time and Stress-Energy Tensor}

\label{p2}

Let us introduce the following spacetime metric%

\begin{equation}
\mathrm{d}s^{2}=-e^{2\Phi\left(  r,{\theta}\right)  }\,\mathrm{d}t^{2}%
+\frac{\mathrm{d}r^{2}}{1-b(r,{\theta})/r}+r^{2}K^{2}(r,{\theta})\left[
\mathrm{d}\theta^{2}+\sin^{2}{\theta}\,\left(  \mathrm{d}\varphi
-\Omega(r,{\theta})\mathrm{d}t\right)  ^{2}\right]  \,\label{metric}%
\end{equation}
representing a stationary, axisymmetric traversable wormhole. The functions
$\Phi\left(  r,{\theta}\right)  $, $b(r,{\theta})$, $K(r,{\theta})$ and
$\Omega(r,{\theta})$ are arbitrary functions of $r$ and ${\theta}$. The range
of the radial coordinate is $\left[  r_{0},+\infty\right)  $, while the
angular variable satisfies ${\theta\in}\left[  0,\pi\right]  $. The function
$\Omega(r,{\theta})$ represents the angular velocity, and $K(r,{\theta})$ is
associated with the proper radial distance. In the limit of $\Omega(r,{\theta
})\rightarrow0$, the line element \eqref{metric} must reproduce the
Morris-Thorne metric \eqref{MT_metric}. We can rearrange the above line
element in the following manner%
\begin{equation}
\mathrm{d}s^{2}=g_{tt}\,\mathrm{d}t^{2}+g_{rr}\mathrm{d}r^{2}+2g_{t\varphi
}\mathrm{d}t\mathrm{d}\varphi+g_{{\theta\theta}}\mathrm{d}\theta
^{2}+g_{\varphi\varphi}\,\mathrm{d}\varphi^{2}\,
\end{equation}
where%
\begin{align}
g_{tt}  &  =-\left(  e^{2\Phi\left(  r,{\theta}\right)  }-r^{2}K^{2}%
(r,{\theta})\sin^{2}{\theta}\Omega^{2}(r,{\theta})\right) \\
g_{rr}  &  =\frac{1}{1-b(r,{\theta})/r} \,\\
g_{t\varphi}  &  =-r^{2}K^{2}(r,{\theta})\sin^{2}{\theta}\Omega(r,{\theta})
\,\\
g_{{\theta\theta}}  &  =r^{2}K^{2}(r,{\theta}) \,\\
g_{\varphi\varphi}  &  =r^{2}K^{2}(r,{\theta})\sin^{2}{\theta}\,.
\end{align}
The ergoregion appears when $g_{tt}$ vanishes, namely%
\begin{equation}
\Phi\left(  r,{\theta}\right)  =\ln\left(  rK(r,{\theta})\Omega(r,{\theta
})\sin{\theta}\right)  {.}\label{ergo}%
\end{equation}
Note also that the discriminant of the metric \eqref{metric} is given by%
\begin{equation}
-g_{tt}g_{\varphi\varphi}+g_{t\varphi}^{2}=e^{2\Phi\left(  r,{\theta}\right)
}r^{2}K^{2}(r,{\theta})\sin^{2}{\theta} \,
\end{equation}
which implies that an event horizon appears when $e^{2\Phi\left(  r,{\theta
}\right)  }=0$. However, such a possibility can be excluded if we aim to
reproduce a TW in the limit of vanishing rotation. In order to reproduce the
correct behavior when the rotation stops, we assume that the redshift function
and the shape function are given by \eqref{PhiCW} and \eqref{b(r)CW} on one
hand, and by \eqref{RFfp} and \eqref{b(r)fp} on the other.

Now that the spacetime has been introduced, we turn our attention to the
anisotropic source tensor. This source includes a radial pressure $p_{r}$, a
transverse pressure $p_{t}$, an energy density $\rho$, and a thermal stress
tensor $\tau$, all described by the following SET
\begin{equation}
T_{\mu\nu}=\left(  \rho+\tau_{\rho}\right)  u_{\mu}u_{\nu}+\left(  p_{r}%
+\tau_{r}\right)  n_{\mu}n_{\nu}+\left(  p_{t}+\tau_{t}\right)  \sigma_{\mu
\nu} \,.\label{SET}%
\end{equation}
The unit timelike vector $u_{\mu}$ is the fluid four-velocity and $n_{\mu}$ is
a unit spacelike vector orthogonal to $u_{\mu}$, implying the relations
$n^{\mu}n_{\mu}=1\,$, $u^{\mu}u_{\mu}=-1$ and $n^{\mu}u_{\mu}=0\,$. The
thermal stress tensor arises from relativistic thermodynamic considerations,
implying that the matter source possesses both mass and heat as local forms of
energy \cite{Hay99}. Moreover, it has been decomposed into an energy component
$\tau_{\rho}\,$, a radial component $\tau_{r}$ and a transverse component
$\tau_{t}\,$. Here,
\begin{equation}
\sigma_{\mu\nu}=g_{\mu\nu}+u_{\mu}u_{\nu}-n_{\mu}n_{\nu}%
\end{equation}
is a projection operator onto a two-surface orthogonal to $u_{\mu}$ and
$n_{\mu}$, i.e.,%
\begin{equation}
u_{\mu}\sigma^{\mu\nu}\mathrm{v}_{\nu}=n_{\mu}\sigma^{\mu\nu}\mathrm{v}_{\nu
}=0\qquad\forall\mathrm{v}_{\nu}\,.\label{orto}%
\end{equation}
The vector $n_{\mu}$ can be written as
\begin{equation}
n_{\mu}=\sqrt{\frac{1}{1-b(r,{\theta})/r}}\left(  0,1,0,0\right)  .
\end{equation}
Keeping in mind the anisotropic structure of the SET, we would like to
introduce rotations. To do this, we define two Killing vectors%
\begin{equation}
k_{t}^{\alpha}=\delta_{t}^{\alpha}\qquad\mathrm{and} \qquad k_{\varphi
}^{\alpha}=\delta_{\varphi}^{\alpha} \,
\end{equation}
whose linear combination allows us to build the four-velocity of the fluid
source%
\begin{equation}
u^{\mu}=\left(  u^{t},0,0,u^{\varphi}\right)  =u^{t}\left(  1,0,0,\Omega
_{0}\right)
\end{equation}
with $\Omega_{0}$ being the angular velocity measured by a distant observer,
defined by%
\begin{equation}
\Omega_{0}=\frac{d\varphi}{dt}=\frac{u^{\varphi}}{u^{t}}.\label{Ome0}%
\end{equation}
Since $u^{\mu}$ is a timelike vector, it must satisfy%
\begin{equation}
u^{\mu}u_{\mu}=\left(  u^{t}\right)  ^{2}\left[  g_{tt}+2g_{t\varphi}%
\Omega_{0}+\Omega_{0}^{2}g_{\varphi\varphi}\right]  =-1\label{Norm}%
\end{equation}
with%
\begin{equation}
g_{tt}+2g_{t\varphi}\Omega_{0}+\Omega_{0}^{2}g_{\varphi\varphi}<0
\,.\label{INorm}%
\end{equation}
In terms of the metric \eqref{metric}, one finds that \eqref{Norm} becomes%
\begin{equation}
\left(  u^{t}\right)  ^{2}=\frac{-1}{\left[  g_{tt}+2g_{t\varphi}\Omega
_{0}+\Omega_{0}^{2}g_{\varphi\varphi}\right]  }=\frac{1}{e^{2\Phi\left(
r,{\theta}\right)  }-r^{2}K^{2}(r,{\theta})\sin^{2}{\theta}\left(
\Omega(r,{\theta})-\Omega_{0}^{2}\right)  ^{2}} \,
\end{equation}
while the inequality \eqref{INorm} becomes
\begin{equation}
\frac{e^{2\Phi\left(  r,{\theta}\right)  }}{r^{2}K^{2}(r,{\theta})\sin
^{2}{\theta}}>\left(  \Omega(r,{\theta})-\Omega_{0}^{2}\right)  ^{2}.
\end{equation}
This means%
\begin{equation}
\Omega_{-}<\Omega(r,{\theta})<\Omega_{+}%
\end{equation}
with
\begin{equation}
\Omega^{\pm}=\Omega_{0}\pm\frac{e^{\Phi\left(  r,{\theta}\right)  }%
}{rK(r,{\theta})\sin{\theta}} \,.
\end{equation}
It is useful to rearrange $u^{\mu}$ in the following manner,
\begin{equation}
u^{\mu}=\frac{e^{-\Phi\left(  r,{\theta}\right)  }}{\left(  1-v^{2}\right)
^{\frac{1}{2}}}\left(  1,0,0,\Omega_{0}\right)  \,
\end{equation}
where%
\begin{equation}
v=rK(r,{\theta})\sin{\theta}\left(  \Omega(r,{\theta})-\Omega_{0}\right)
\,e^{-\Phi\left(  r,{\theta}\right)  }%
\end{equation}
is the proper velocity of the matter with respect to a zero angular momentum
observer. With the help of the line element \eqref{metric}, we can compute the
covariant form of $u_{\mu}\,$, which is given by%
\begin{align*}
u_{t}  &  =\frac{e^{-\Phi\left(  r,{\theta}\right)  }}{\left(  1-v^{2}\right)
^{\frac{1}{2}}}\left(  g_{tt}+\Omega_{0}g_{t\varphi}\right) \\
u_{\varphi}  &  =\frac{e^{-\Phi\left(  r,{\theta}\right)  }}{\left(
1-v^{2}\right)  ^{\frac{1}{2}}}\left(  g_{t\varphi}+\Omega_{0}g_{\varphi
\varphi}\right)  .
\end{align*}
With this information, we can write the components of the SET \eqref{SET} as
follows:
\begin{align}
T_{tt}  &  =\left(  \rho+\tau_{\rho}\right)  u_{t}u_{t}+\left(  p_{t}+\tau
_{t}\right)  \left(  g_{tt}+u_{t}u_{t}\right)  =\left(  \rho+\tau_{\rho}%
+p_{t}+\tau_{t}\right)  u_{t}u_{t}+\left(  p_{t}+\tau_{t}\right)
g_{tt}\label{Ttt}\\
T_{rr}  &  =\left(  p_{r}+\tau_{r}\right)  n_{r}n_{r}=\left(  p_{r}%
+\tau_{r}\right)  g_{rr}\label{Trr}\\
T_{{\theta\theta}}  &  =\left(  p_{t}+\tau_{t}\right)  \sigma_{{\theta\theta}%
}=\left(  p_{t}+\tau_{t}\right)  g_{{\theta\theta}}\label{Tthth}\\
T_{\varphi\varphi}  &  =\left(  \rho+\tau_{\rho}\right)  u_{\varphi}%
u_{\varphi}+\left(  p_{t}+\tau_{t}\right)  g_{\varphi\varphi}+\left(
p_{t}+\tau_{t}\right)  u_{\varphi}u_{\varphi}=\left(  \rho+\tau_{\rho}%
+p_{t}+\tau_{t}\right)  u_{\varphi}u_{\varphi}+\left(  p_{t}+\tau_{t}\right)
g_{\varphi\varphi}\label{Tpp}\\
T_{t\varphi}  &  =T_{\varphi t}=\left(  \rho+\tau_{\rho}\right)
u_{t}u_{\varphi}+\left(  p_{t}+\tau_{t}\right)  \sigma_{t\varphi}=\left(
\rho+\tau_{\rho}+p_{t}+\tau_{t}\right)  u_{t}u_{\varphi}+\left(  p_{t}%
+\tau_{t}\right)  g_{t\varphi} \,.\label{Ttp}%
\end{align}
For the specific case under investigation, the property $\rho+p_{t}=0$ holds.
Therefore, the previous components \eqref{Ttt}-\eqref{Ttp} can be rearranged
as
\begin{align}
T_{tt}  &  =\left(  \tau_{\rho}+\tau_{t}\right)  u_{t}u_{t}+\left(  p_{t}%
+\tau_{t}\right)  g_{tt}\label{TttC}\\
T_{rr}  &  =\left(  p_{r}+\tau_{r}\right)  g_{rr}\label{TrrC}\\
T_{{\theta\theta}}  &  =\left(  p_{t}+\tau_{t}\right)  g_{{\theta\theta}%
}\label{TththC}\\
T_{\varphi\varphi}  &  =\left(  \tau_{\rho}+\tau_{t}\right)  u_{\varphi
}u_{\varphi}+\left(  p_{t}+\tau_{t}\right)  g_{\varphi\varphi}\label{TppC}\\
T_{t\varphi}  &  =T_{\varphi t}=\left(  \tau_{\rho}+\tau_{t}\right)
u_{t}u_{\varphi}+\left(  p_{t}+\tau_{t}\right)  g_{t\varphi\varphi}
\,\label{TtpC}%
\end{align}
where $\tau_{\rho}$, $\tau_{t}$ and $\tau_{r}$ need to be determined. It is
straightforward to see that $T_{rr}$ and $T_{{\theta\theta}}$ are not affected
by the rotation. It is also convenient to introduce a particular reference
frame, called ZAMO (Zero Angular Momentum Observer), in which $u_{\varphi
}=u^{t}g_{t\varphi}+u^{\varphi}g_{\varphi\varphi}=0\,$. This means that
\begin{align}
\Omega_{0}=\frac{u^{\varphi}}{u^{t}}=-\frac{g_{t\varphi}}{g_{\varphi\varphi}%
}=\Omega(r,{\theta})\qquad\Longrightarrow\qquad v=0 \,.
\end{align}
Moreover, we find%
\begin{equation}
u_{t}=\frac{e^{-\Phi\left(  r,{\theta}\right)  }}{\left(  1-v^{2}\right)
^{\frac{1}{2}}}\left(  g_{tt}+\Omega_{0}g_{t\varphi}\right)  =-e^{\Phi\left(
r,{\theta}\right)  } \,,
\end{equation}
as expected for a ZAMO. With these ingredients, the previous SET can be
written as
\begin{align}
T_{tt}  &  =p_{t}g_{tt}+\tau_{t}\left(  g_{tt}+e^{2\Phi\left(  r,{\theta
}\right)  }\right)  =p_{t}g_{tt}+\tau_{t}g_{\varphi\varphi}\Omega
^{2}(r,{\theta})\label{Ttt1}\\
T_{rr}  &  =p_{r}n_{r}n_{r}=p_{r}g_{rr}\label{Trr1}\\
T_{{\theta\theta}}  &  =p_{t}\sigma_{{\theta\theta}}=p_{t}g_{{\theta\theta}%
}\label{Tthth1}\\
T_{\varphi\varphi}  &  =p_{t}g_{\varphi\varphi}\label{Tpp1}\\
T_{t\varphi}  &  =T_{\varphi t}=p_{t}g_{t\varphi}=-p_{t}\Omega(r,{\theta
})g_{\varphi\varphi} \,.\label{Ttp1}%
\end{align}

\section{Einstein Field Equations for a rotating Traversable Wormhole}

\label{p3}

One crucial aspect of rotations is the emergence of a constraint related to
the field equation $G_{r\theta}\,$, which must vanish \cite{PBH,Kuhfittig}.
This implies that some restrictions are necessary. To this purpose, we assume
that
\begin{align}
b\left(  r,\theta\right)   &  \rightarrow b\left(  r\right) \\
K(r,{\theta})  &  \rightarrow1\\
\Phi\left(  r,\theta\right)   &  \rightarrow\Phi\left(  r\right)  .
\end{align}
With this choice, the aforementioned constraint equation reduces to
\begin{equation}
G_{r\theta}=\frac{r^{2}}{2}\sin^{2}\left(  \theta\right)  e^{-2\Phi\left(
r\right)  }\frac{\partial\Omega\left(  r,\theta\right)  }{\partial\theta}%
\frac{\partial\Omega\left(  r,\theta\right)  }{\partial r}=0
\end{equation}
and it can be satisfied if
\begin{equation}
\Omega\left(  r,\theta\right)  \rightarrow\Omega\left(  r\right) \label{or}%
\end{equation}
or
\begin{equation}
\Omega\left(  r,\theta\right)  \rightarrow\Omega\left(  \theta\right)  .
\end{equation}
In this paper, we will adopt the choice \eqref{or}. The restriction of
$b(r,{\theta})$ to $b(r)$ is also dictated by the presence of a singularity in
the scalar curvature $R\,$, which proportional to%
\begin{equation}
R\sim\frac{\partial_{\theta}b(r,{\theta})}{\left(  r-r_{0}\right)  ^{2}} \,.
\end{equation}
This singularity can be eliminated by imposing that $\partial_{\theta
}b(r,{\theta})=0$ \cite{Teo}. We can observe that the SET components
\eqref{TrrC}-\eqref{TppC} are not affected by the rotation. In the next
subsection, we will examine the rotating structure of a Casimir Wormhole. The
components of the SET \eqref{TttC}-\eqref{TtpC} suggest considering firstly
those equations that are independent of the rotation. We will consider two
models regarding the Casimir plates:

\begin{description}
\item[a)] Rotating Casimir Wormhole with radially varying plates

\item[b)] Rotating Casimir Wormhole with fixed plates.
\end{description}

We begin by considering the case \textbf{a)}.

\subsection{Rotating Casimir Wormhole with radially varying plates}

We begin by considering the EFE that does not include angular velocity in the
SET. To this end, the first equation we will examine is $G_{rr}=\kappa
T_{rr}\,$, which is described by%
\begin{equation}
\frac{r-b\left(  r\right)  }{4r^{2}}\left(  r^{3}\left(  \Omega^{\prime
}\left(  r\right)  \right)  ^{2}\sin^{2}\theta e^{-2\Phi\left(  r\right)
}+8\Phi^{\prime}\left(  r\right)  \right)  -\frac{b\left(  r\right)  }{r^{3}%
}+\frac{3r_{1}^{2}}{r^{4}}-\kappa\tau_{r}\left(  r\right)  =0 \,.\label{EFErr}%
\end{equation}
At the throat, the previous equation reduces to
\begin{equation}
-r_{0}^{2}+3r_{1}^{2}=r_{0}^{4}\kappa\tau_{r}\left(  r_{0}\right)
\end{equation}
upon using the throat condition $b\left(  r_{0}\right)  =r_{0}\,$. It is
immediate to see that a solution is%
\begin{equation}
r_{0}=\sqrt{3} r_{1} \qquad\mathrm{and} \qquad\tau_{r}\left(  r_{0}\right)  =0
\,,\label{s}%
\end{equation}
which is characteristic of a Casimir Wormhole. To extend the solution outside
the throat, we substitute \eqref{PhiCW} and \eqref{b(r)CW} into the EFE
\eqref{EFErr} to obtain
\begin{equation}
\frac{\sin^{2}\left(  \theta\right)  r^{2}\left(  r-r_{0}\right)  \left(
3r+r_{0}\right)  ^{3}\left(  \Omega^{\prime}\left(  r\right)  \right)  ^{2}%
}{108\kappa}=\tau_{r}\left(  r\right)  .
\end{equation}
Of course, in the vicinity of the throat, the left-hand side vanishes, and a
solution emerges if we assume that $\tau_{r}\left(  r\right)  $ also vanishes.
However, in this approximation, $\Omega\left(  r\right)  $ remains
undetermined. Another solution with $\tau_{r}\left(  r\right)  =0$ arises when
$\Omega\left(  r\right)  =\Omega\,$, with $\Omega$ being constant. This time,
the solution is valid for all $r\in\left[  r_{0},+\infty\right)  $.

The next equation we will consider is the EFE $G_{{\theta\theta}}=\kappa
T_{{\theta\theta}}$, which can be reduced to%
\begin{equation}
\frac{\left(  9r+r_{0}\right)  r_{0}^{2}}{3r^{4}\left(  3r+r_{0}\right)
}=\kappa\tau_{t}\left(  r\right)  +\frac{r_{1}^{2}}{r^{4}} \,.\label{EFEthth}%
\end{equation}
To obtain \eqref{EFEthth}, we have plugged \eqref{PhiCW}, \eqref{b(r)CW} and
$\Omega\left(  r\right)  =\Omega$ into \eqref{GTthth}, and we have also used
the relationship $r_{0}=\sqrt{3}r_{1}\,$. Isolating $\tau_{t}\left(  r\right)
$, one finds%
\begin{equation}
\tau_{t}\left(  r\right)  =\frac{2r_{0}^{2}}{\kappa r^{3}\left(
3r+r_{0}\right)  } \,.\label{taut}%
\end{equation}
The next equation we are about to examine is the EFE $G_{\phi\phi}=\kappa
T_{\phi\phi}\,$, which can be reduced to
\begin{align}
&  r_{0}^{2}\left(  9r+r_{0}\right) \nonumber\\
&  =-\frac{\left(  \left(  \Omega-\Omega_{0}\right)  ^{2}\left(  3\tau_{\rho
}\left(  r\right)  \kappa\,r^{4}-r_{0}^{2}\right)  \left(  3r+r_{0}\right)
^{3}\sin^{2}\left(  \theta\right)  +9r_{0}^{2}\left(  9r+r_{0}\right)
\right)  }{\left(  \left(  \Omega-\Omega_{0}\right)  \left(  3r+r_{0}\right)
\sin\left(  \theta\right)  \right)  ^{2}-9} \,.\label{EFEpp}%
\end{align}
To obtain \eqref{EFEpp}, we have plugged \eqref{PhiCW}, \eqref{b(r)CW},
\eqref{taut}, $\Omega\left(  r\right)  =\Omega$ and $r_{0}=\sqrt{3}r_{1}$ into
\eqref{GTpp}. A solution of this equation is%
\begin{equation}
\tau_{\rho}\left(  r\right)  =-\frac{2r_{0}^{2}}{\kappa r^{3}\left(
3r+r_{0}\right)  } \,.\label{taurho}%
\end{equation}
Alternatively, one can observe that if we assume%
\begin{equation}
\tau_{\rho}\left(  r\right)  =\frac{r_{0}^{2}}{3\kappa r^{4}}%
\,,\label{taurho1}%
\end{equation}
Eq.\eqref{EFEpp} becomes%
\begin{equation}
\left(  \Omega-\Omega_{0}\right)  \left(  3r+r_{0}\right)  \sin\left(
\theta\right)  ^{2}=0 \,
\end{equation}
and is satisfied only if $\Omega=\Omega_{0}\,$, that is, in a ZAMO frame. The
last two EFE that need to be examined are $G_{{tt}}=\kappa T_{tt}$ and
$G_{t\phi}=\kappa T_{t\phi}\,$. We begin with $G_{t\phi}=\kappa T_{t\phi}\,$.
From \eqref{GTtp} and with the help of \eqref{PhiCW}, \eqref{b(r)CW},
\eqref{taut}, $\Omega\left(  r\right)  =\Omega\,$, and $r_{0}=\sqrt{3}r_{1}%
\,$, one can write%
\begin{equation}
-A\left(  \Omega,r\right)  \tau_{\rho}\left(  r\right)  -B\left(
\Omega,r\right)  =0\label{EFEtp}%
\end{equation}
where%
\begin{align}
A\left(  \Omega,r\right)   &  =\frac{\left(  -1+\left(  \Omega-\Omega
_{0}\right)  \Omega\left(  r+\frac{r_{0}}{3}\right)  ^{2}\left(  \sin
^{2}\left(  \theta\right)  \right)  \right)  r^{2}\left(  \Omega-\Omega
_{0}\right)  \left(  \sin^{2}\left(  \theta\right)  \right)  }{\left(  \left(
\Omega-\Omega_{0}\right)  \left(  r+\frac{r_{0}}{3}\right)  \sin\left(
\theta\right)  \right)  ^{2}-1}\\
B\left(  \Omega,r\right)   &  =\frac{2\left(  \sin^{2}\left(  \theta\right)
\right)  \left(  -1+\left(  \Omega-\Omega_{0}\right)  \Omega\left(
r+\frac{r_{0}}{3}\right)  ^{2}\left(  \sin^{2}\left(  \theta\right)  \right)
\right)  \left(  \Omega-\Omega_{0}\right)  r_{0}^{2}}{3r\left(  \left(
\left(  \Omega-\Omega_{0}\right)  \left(  r+\frac{r_{0}}{3}\right)
\sin\left(  \theta\right)  \right)  ^{2}-1\right)  \left(  r+\frac{r_{0}}%
{3}\right)  \kappa} \,.
\end{align}
Solving for $\tau_{\rho}\left(  r\right)  $, one finds \eqref{taurho}. This
implies that the solution \eqref{taurho1} can be discarded. Finally, we must
examine the EFE $G_{{tt}}=\kappa T_{tt}\,$, as described by \eqref{GTtt}. Even
in this last EFE, with the help of \eqref{PhiCW}, \eqref{b(r)CW},
\eqref{taut}, $\Omega\left(  r\right)  =\Omega\,$, and $r_{0}=\sqrt{3}r_{1}%
\,$, one can write, for $\Omega\neq\Omega_{0}\,$,%
\begin{equation}
\frac{C\left(  \Omega,r\right)  }{F\left(  \Omega,r\right)  }\tau_{\rho
}\left(  r\right)  +\frac{r_{0}^{2}\left(  D\left(  \Omega,r\right)  -E\left(
\Omega,r\right)  \right)  }{3r^{2}\kappa F\left(  \Omega,r\right)  }%
=\frac{9r_{0}^{2}G\left(  \Omega,r\right)  }{\kappa r^{2}\left(
3r+r_{0}\right)  ^{2}}\label{EFEtt}%
\end{equation}
where we have defined%
\begin{align}
C\left(  \Omega,r\right)   &  =r^{2}\left(  \left(  3r+r_{0}\right)
^{2}\left(  \Omega-\Omega_{0}\right)  \Omega\left(  \sin^{2}\left(
\theta\right)  \right)  -9\right)  ^{2}\\
D\left(  \Omega,r\right)   &  =81+\left(  3r+r_{0}\right)  ^{4}\left(
\Omega-\Omega_{0}\right)  ^{2}\Omega^{2}\sin^{4}\left(  \theta\right) \\
E\left(  \Omega,r\right)   &  =9\left(  3r+r_{0}\right)  \left[  \left(
3r+r_{0}\right)  \left(  2\left(  \Omega-\Omega_{0}\right)  ^{2}+\Omega
_{0}\left(  2\Omega+\Omega_{0}\right)  \right)  -2r_{0}\Omega_{0}^{2}\right]
\sin^{2}\left(  \theta\right) \\
F\left(  \Omega,r\right)   &  =\left(  3r+r_{0}\right)  ^{2}\left(  \left(
3r+r_{0}\right)  ^{2}\left(  \Omega-\Omega_{0}\right)  ^{2}\left(  \sin
^{2}\left(  \theta\right)  \right)  -9\right) \\
G\left(  \Omega,r\right)   &  =\frac{1}{27}\left(  \Omega^{2}\left(
9r+r_{0}\right)  \left(  3r+r_{0}\right)  \sin^{2}\left(  \theta\right)
-9\right)  .
\end{align}
Eq.~\eqref{EFEtt} admits no solution for $\Omega\neq\Omega_{0}$. Nevertheless,
for the special case $\Omega=\Omega_{0}$ (ZAMO), Eq.~\eqref{EFEpp} reduces to
an identity, and $\tau_{\rho}\left(  r\right)  $ can be no more determined.
The same occurs for Eq.\eqref{EFEtp}. Therefore, in a ZAMO frame, Eq.
\eqref{EFEtt} becomes%
\begin{align}
\frac{r^{2}\tau_{\rho}\left(  r\right)  }{\left(  3r+r_{0}\right)  ^{2}}%
+\frac{r_{0}^{2}\left(  9-\left(  \left(  3r+r_{0}\right)  \left(
9r+r_{0}\right)  \Omega^{2}\sin^{2}\left(  \theta\right)  \right)  \right)
}{27r^{2}\kappa\left(  3r+r_{0}\right)  ^{2}} =\frac{r_{0}^{2}\left(
9-\Omega^{2}\left(  9r+r_{0}\right)  \left(  3r+r_{0}\right)  \sin^{2}\left(
\theta\right)  \right)  }{27\kappa r^{2}\left(  3r+r_{0}\right)  ^{2}}%
\end{align}
and the only solution is
\begin{equation}
\tau_{\rho}\left(  r\right)  =0 \,.\label{taurho0}%
\end{equation}
Apparently, a contradiction arises between \eqref{taurho} and \eqref{taurho0}.
However, this is not the case, as we are compelled to set $\Omega=\Omega
_{0}\,$, which means that the only consistent solution is the vanishing of
$\tau_{\rho}\left(  r\right)  \, $. Even though the constant $\Omega$ is a
solution of the EFE, it has the unpleasant feature of not vanishing at large
distances. This implies that a dragging effect will be present even at
infinity. To address this, we observe that, by considering the definition of
an ergosurface, and to avoid a change in the signature, we can assume to
remain outside the ergoregion defined by \eqref{ergo}, namely
\begin{equation}
\frac{3}{\left(  3r+r_{0}\right)  \sin{\theta}}>\Omega\,.\label{Ome}%
\end{equation}
Due to the inequality \eqref{Ome}, one can argue that the farther the distance
from the throat, the smaller the value of $\Omega$ becomes, even though the
decrease in the dragging velocity is not very rapid. Of course, this argument
can only be applied if we are far from the values ${\theta=0}$ and
${\theta=\pi}$. Note that the inequality also represents an upper bound when
we are at the throat. Indeed, we have
\begin{equation}
\frac{3}{4r_{0}\sin{\theta}}=\frac{\sqrt{3}}{4r_{1}\sin{\theta}}%
>\Omega\label{UB}%
\end{equation}
by using \eqref{s}. Due to the inequalities \eqref{Ome} and \eqref{UB}, a
better strategy is needed to describe the vanishing of the rotation when
$r>r_{0}\,$. To this end, we propose the following profile
\begin{equation}
\Omega\left(  r\right)  =\Omega\exp\left(  -\mu\left(  r-r_{0}\right)
\right)  .\label{DO}%
\end{equation}
Such a profile can be introduced at the cost of considering a non-vanishing
$\tau_{r}\left(  r\right)  $. Indeed, Eq.~\eqref{EFErr} is satisfied provided
that one considers
\begin{equation}
\tau_{r}\left(  r\right)  =\frac{\left(  3r+r_{0}\right)  ^{3}\left(  \sin
^{2}\left(  \theta\right)  \right)  \left(  r-r_{0}\right)  \Omega^{2}\mu
^{2}e^{-2\mu\left(  r-r_{0}\right)  }}{108\kappa\,r^{2}} \,.
\end{equation}
As we can see, $\tau_{r}\left(  r\right)  $ vanishes at the throat, as in the
pure constant $\Omega$ case, and is highly suppressed for $r>r_{0}\,$. When
$\Omega( r) $ is plugged into \eqref{EFEthth}, we obtain the following
modification for $\tau_{t}\left(  r\right)  $:
\begin{equation}
\tau_{t}\left(  r\right)  =-\frac{3\Omega^{2}\left(  3r+r_{0}\right)  ^{3}%
\mu^{2}\left(  r-r_{0}\right)  \left(  \sin^{2}\left(  \theta\right)  \right)
e^{-2\mu\left(  r-r_{0}\right)  }}{108\kappa r^{2}}+\frac{2r_{0}^{2}}{\kappa
r^{3}\left(  3r+r_{0}\right)  }\,.\label{taut1}%
\end{equation}
Close to the throat, the first term of \eqref{taut1} can be neglected compared
to the second term, and the form of \eqref{taut} is recovered. For $r>r_{0}%
\,$, the exponential suppresses the first term, and the behavior of
\eqref{taut} persists. By plugging $\tau_{t}\left(  r\right)  $ as described
by \eqref{taut1} and \eqref{DO} into the EFE \eqref{EFEtt}, one finds that
\eqref{taurho0} remains valid. Moreover, Eq.~\eqref{taurho0} is satisfied both
for large values of $\mu$ and $r$, as well as for $\mu=0$ and close to the
throat. Regarding the EFE $G_{\phi\phi}=\kappa T_{\phi\phi}$ and the EFE
$G_{t\phi}=\kappa T_{t\phi}\,$, we note that for $\mu=0$ and close to the
throat, both are satisfied, while for large values of $\mu$ and $r\,$, they
vanish, but with different numerical coefficients.

It remains to check the violation of the Null Energy Condition (NEC). We
recall that the NEC is violated if
\begin{equation}
\rho\left(  r\right)  +p_{r}\left(  r\right)  \leq0 \,.
\end{equation}
Since the violation of the NEC is relevant close to the throat, it is not
necessary to introduce $\tau_{r}\left(  r\right)  $ and $\Omega\left(
r\right)  =\Omega$. Therefore, in this context, we can write%
\begin{equation}
G_{\mu\nu}u^{\mu}u^{\nu}+G_{\mu\nu}n^{\mu}n^{\nu}=\kappa\left(  T_{\mu\nu
}u^{\mu}u^{\nu}+T_{\mu\nu}n^{\mu}n^{\nu}\right)  =\kappa\left(  \rho\left(
r\right)  +p_{r}\left(  r\right)  \right)
\end{equation}
or, in other words,
\begin{equation}
-\frac{2r_{0}^{2}\left(  \left(  3r+r_{0}\right)  \left(  9r+2r_{0}\right)
\left(  \Omega-\Omega_{0}\right)  ^{2}\sin^{2}\theta-18\right)  }{r^{4}\left(
\left(  3r+r_{0}\right)  ^{2}\left(  \Omega-\Omega_{0}\right)  ^{2}\sin
^{2}\theta-9\right)  }=\kappa\left(  \rho\left(  r\right)  +p_{r}\left(
r\right)  \right)  .\label{rhopr}%
\end{equation}
When $\Omega=\Omega_{0}\,$, we get
\begin{equation}
-\frac{4r_{0}^{2}}{3r^{4}}=\kappa\left(  \rho\left(  r\right)  +p_{r}\left(
r\right)  \right)  \, ,
\end{equation}
as expected. Nevertheless, if we approach the throat before considering a ZAMO
in \eqref{rhopr}, we obtain
\begin{equation}
\frac{4\left(  9-22r_{0}^{2}\left(  \Omega-\Omega_{0}\right)  ^{2}\sin
^{2}\theta\right)  }{r_{0}^{2}\left(  16r_{0}^{2}\left(  \Omega-\Omega
_{0}\right)  ^{2}\sin^{2}\theta-9\right)  }=\kappa\left(  \rho\left(
r_{0}\right)  +p_{r}\left(  r_{0}\right)  \right)  .\label{rhopr0}%
\end{equation}
It is straightforward to see that there are two values of $\Omega$ such that
$\rho\left(  r_{0}\right)  +p_{r}\left(  r_{0}\right)  =0\,$. These are
\begin{equation}
\Omega_{1,2}=\Omega_{0}\pm\frac{3\sqrt{22}}{22r_{0}\sin\theta} \,.\label{N1}%
\end{equation}
On the other hand, the denominator of \eqref{rhopr0} vanishes for%
\begin{equation}
\Omega_{3,4}=\Omega_{0}\pm\frac{3}{4r_{0}\sin\theta} \,.
\end{equation}
This means that for values of $\Omega$ such that%
\begin{align}
\Omega_{3}  &  >\Omega>\Omega_{1}\\
\Omega_{2}  &  >\Omega>\Omega_{4}\,,
\end{align}
the NEC is not violated. Moreover, from \eqref{rhopr0}, one finds in a ZAMO
frame that
\begin{equation}
-\frac{4}{r_{0}^{2}}=\kappa\left(  \rho\left(  r_{0}\right)  +p_{r}\left(
r_{0}\right)  \right)  .
\end{equation}
In other words, it appears that%
\begin{equation}
\lim_{r \rightarrow r_{0}}\lim_{\Omega\rightarrow\Omega_{0}}\kappa\left(
\rho\left(  r\right)  +p_{r}\left(  r\right)  \right)  \neq\lim_{\Omega
\rightarrow\Omega_{0}}\lim_{r \rightarrow r_{0}}\kappa\left(  \rho\left(
r\right)  +p_{r}\left(  r\right)  \right)  \, .\label{NC}%
\end{equation}

\subsection{Rotating Casimir Wormhole with fixed plates}

In this subsection, we consider the profile given by \eqref{b(r)fp} and aim to
apply the same method used for the SET in \eqref{CSET}, but now with the SET
described by \eqref{Casimir}. The EFE are formally the same as in the previous
subsection, with the only modification being the change of the plates'
separation from $r$ to $d$ on the SET components. The first EFE we wish to
examine, when rotations come into play, is the equation $G_{rr}=\kappa
T_{rr}\,$, which reads
\begin{equation}
\frac{r-b\left(  r\right)  }{4r^{2}}\left(  r^{3}\left(  \Omega^{\prime
}\left(  r\right)  \right)  ^{2}\sin^{2}\theta e^{-2\Phi\left(  r\right)
}+8\Phi^{\prime}\left(  r\right)  \right)  -\frac{b\left(  r\right)  }{r^{3}%
}+\frac{3r_{1}^{2}}{d^{4}}-\kappa\tau_{r}\left(  r\right)  =0
\,.\label{EFErrd}%
\end{equation}
Since the condition $b\left(  r_{0}\right)  =r_{0}$ must be satisfied, we
observe that, at the throat, the previous equation simplifies to
\begin{equation}
-\frac{1}{r_{0}^{2}}+\frac{3r_{1}^{2}}{d^{4}}-\kappa\tau_{r}\left(
r_{0}\right)  =0 \,.
\end{equation}
Its solution is given by \eqref{r0d}, along with the assumption that $\tau
_{r}\left(  r_{0}\right)  =0\,$. Isolating $\tau_{r}\left(  r\right)  $ from
\eqref{EFErrd}, we get
\begin{equation}
\tau_{r}\left(  r\right)  =\frac{\left(  r^{7}+9r^{5}r_{0}^{2}-10r^{4}%
r_{0}^{3}\right)  \sin^{2}\left(  \theta\right)  \left(  \Omega^{\prime
}\left(  r\right)  \right)  ^{2}+40r^{3}-40r_{0}^{3}}{36r_{0}^{2}\kappa r^{3}}
\,
\end{equation}
where we have used \eqref{RFfp}, \eqref{b(r)fp} and \eqref{r0d}. Because of
the boundary \eqref{rbar}, we are forced to remain in the vicinity of the
throat. Thus, we can write
\begin{equation}
\tau_{r}\left(  r\right)  \simeq\frac{10\left(  r^{3}-r_{0}^{3}\right)
}{9r_{0}^{2}\kappa r^{3}} \,.
\end{equation}
Note that $\Omega\left(  r\right)  $ still remains undetermined. Therefore, we
next consider the EFE $G_{\theta\theta}=\kappa T_{\theta\theta}\,$. Isolating
$\tau_{t}\left(  r\right)  $ from \eqref{GTthth}, one gets
\begin{equation}
\tau_{t}\left(  r\right)  =\frac{\sin^{2}\left(  \theta\right)  \left(
\Omega^{\prime}\left(  r\right)  \right)  ^{2}\left(  10r^{4}r_{0}^{3}%
-r^{7}-9r^{5}r_{0}^{2}\right)  -8r^{3}+20r_{0}^{3}}{36r_{0}^{2}\kappa\,r^{3}}
\,.\label{tautfp}%
\end{equation}
Once again, in proximity of the throat, the equation is independent of
$\Omega\left(  r\right)  $ and reduces to%
\begin{equation}
\tau_{t}\left(  r\right)  \simeq\frac{5r_{0}^{3}-2r^{3}}{9\kappa r_{0}%
^{2}\,r^{3}}.
\end{equation}
As a result, we will evaluate $G_{\phi\phi}=\kappa T_{\phi\phi}\,$. From
\eqref{GTpp}, it is easy to show that, in the vicinity of the throat,
$G_{\phi\phi}$ reduces to
\begin{equation}
G_{\phi\phi}\simeq\frac{\left(  r\left(  \Phi^{\prime}\left(  r\right)
\right)  +1\right)  }{2\kappa r}\left(  b\left(  r\right)  -\left(  b^{\prime
}\left(  r\right)  \right)  r\right)  =\frac{r^{3}+5r_{0}^{3}}{9r_{0}%
^{2}\kappa r}%
\end{equation}
while $T_{\phi\phi}$ becomes%
\begin{equation}
T_{\phi\phi}=\frac{r^{3}+5r_{0}^{3}}{9r_{0}^{2}\kappa r}\sin^{2}\left(
\theta\right)  +\frac{\left(  20r_{0}^{3}-8r^{3}\right)  \left(  \Omega\left(
r\right)  -\Omega_{0}\right)  ^{2}r\sin^{4}\left(  \theta\right)  }%
{36r_{0}^{2}\kappa\left(  1-r^{2}\left(  \sin^{2}\left(  \theta\right)
\right)  \left(  \Omega\left(  r\right)  -\Omega_{0}\right)  ^{2}\right)  }%
\end{equation}
Still $\Omega\left(  r\right)  $ remains undetermined. However, if we adopt
the ZAMO frame with the additional assumption that $\Omega\left(  r\right)
=\Omega$ with $\Omega$ constant, then $G_{\phi\phi}=\kappa T_{\phi\phi}$ is satisfied.

It remains to evaluate $G_{tt}=\kappa T_{tt}$ and $G_{t\phi}=\kappa T_{t\phi}
$. We begin with $G_{tt}=\kappa T_{tt}$. By examining $G_{tt}$ from
\eqref{GTtt}. one finds, in vicinity of the throat, that
\begin{align}
&  G_{tt}\simeq\frac{2b^{\prime}\left(  r\right)  e^{2\Phi\left(  r\right)
}-r\Omega\left(  r\right)  ^{2}\sin^{2}\left(  \theta\right)  \left(
b^{\prime}\left(  r\right)  r-b\left(  r\right)  \right)  }{2r^{2}}-\frac
{\sin^{2}\left(  \theta\right)  \Omega\left(  r\right)  ^{2}\left(  b^{\prime
}\left(  r\right)  r+b\left(  r\right)  -2r\right)  \Phi^{\prime}\left(
r\right)  }{2}\nonumber\\
&  +\Omega^{\prime}\left(  r\right)  \sin^{2}\left(  \theta\right)
\Omega\left(  r\right)  \left(  \frac{b^{\prime}\left(  r\right)  r}%
{2}-4r+\frac{7b\left(  r\right)  }{2}\right)  .\label{GTttG}%
\end{align}
By plugging \eqref{RFfp}, \eqref{b(r)fp} and \eqref{r0d} into \eqref{GTttG},
we obtain
\begin{equation}
G_{tt}\simeq\frac{2b^{\prime}\left(  r\right)  -r\Omega^{2}\sin^{2}\left(
\theta\right)  \left(  b^{\prime}\left(  r\right)  r-b\left(  r\right)
\right)  }{2r^{2}}=\frac{\sin^{2}\left(  \theta\right)  \Omega^{2}\left(
r^{3}+5r_{0}^{3}\right)  -3r}{9r_{0}^{2}\kappa r}%
\end{equation}
where we have also used that $\Omega\left(  r\right)  =\Omega\,$. On the other
hand, $T_{tt}$ becomes
\begin{align}
&  T_{tt}\simeq\left(  \frac{r_{1}^{2}}{\kappa\,d^{4}}+\tau_{t}\left(
r\right)  \right)  \left(  r^{2}\Omega^{2}\sin^{2}\left(  \theta\right)
-1\right)  +\frac{\tau_{t}\left(  r\right)  \left(  r^{2}\sin^{2}\left(
\theta\right)  \left(  \Omega^{2}-\Omega\Omega_{0}\right)  -1\right)  ^{2}%
}{1-r^{2}\sin^{2}\left(  \theta\right)  \left(  \Omega-\Omega_{0}\right)
^{2}}\nonumber\\
&  =\frac{3r^{5}\Omega^{2}\left(  \Omega-\Omega_{0}\right)  ^{2}\sin
^{4}\left(  \theta\right)  +\left(  \left(  -6\Omega^{2}+6\Omega\Omega
_{0}-\Omega_{0}^{2}\right)  r^{3}-5\Omega_{0}^{2}r_{0}^{3}\right)  \sin
^{2}\left(  \theta\right)  +3r}{9r_{0}^{2}r\left(  r^{2}\sin^{2}\left(
\theta\right)  \left(  \Omega-\Omega_{0}\right)  ^{2}-1\right)  \kappa}%
\end{align}
where we have set $\tau_{\rho}\left(  r\right)  =0$ and used \eqref{tautfp}.
Even in this case, the ZAMO frame significantly simplifies the expression for
$T_{tt}\,$. Indeed, we find%
\begin{equation}
T_{tt}=\frac{\Omega^{2}\left(  r^{3}+5r_{0}^{3}\right)  \sin^{2}\left(
\theta\right)  -3r}{9\kappa r_{0}^{2}r}%
\end{equation}
and, as a result, the EFE $G_{tt}=\kappa T_{tt}$ is satisfied.

The last equation to examine is $G_{t\phi}=\kappa T_{t\phi}\,$. As with the
other equations, we write $G_{t\phi}$ in proximity of the throat as
\begin{equation}
G_{t\phi}\simeq-\frac{\Omega}{2\kappa r}\left(  \left(  b^{\prime}\left(
r\right)  \right)  r-b\left(  r\right)  \right)  =\frac{\sin^{2}\left(
\theta\right)  \Omega\left(  r^{3}+5r_{0}^{3}\right)  \sin^{2}\left(
\theta\right)  }{9\kappa r_{0}^{2}r}%
\end{equation}
where we have used that $\Omega\left(  r\right)  =\Omega$ together with
\eqref{RFfp}, \eqref{b(r)fp} and \eqref{r0d}. The component $T_{t\phi}$, in
the same approximation, becomes
\begin{equation}
T_{t\phi}\simeq-\frac{r^{3}+5r_{0}^{3}}{9r_{0}^{2}\kappa r}\Omega\sin
^{2}\left(  \theta\right)
\end{equation}
where we have also used the ZAMO frame and Eq.~\eqref{tautfp}. Therefore, even
the last EFE is satisfied, at least near the throat. Since the boundary
\eqref{rbar} is very close to the throat, the approximation we used is
consistent. However, for this model, we still need to check the NEC violation.
In this context, we can write
\begin{equation}
G_{\mu\nu}u^{\mu}u^{\nu}+G_{\mu\nu}n^{\mu}n^{\nu}=\kappa\left(  T_{\mu\nu
}u^{\mu}u^{\nu}+T_{\mu\nu}n^{\mu}n^{\nu}\right)  =\kappa\left(  \rho\left(
r\right)  +p_{r}\left(  r\right)  +\tau_{r}\left(  r\right)  \right)
\end{equation}
or, in other words,%
\begin{equation}
\frac{2r^{3}+10r_{0}^{3}-15r^{2}r_{0}^{3}\left(  \Omega-\Omega_{0}\right)
^{2}\sin^{2}\left(  \theta\right)  }{9r_{0}^{2}\left(  \left(  \Omega
-\Omega_{0}\right)  ^{2}r^{2}\sin^{2}\left(  \theta\right)  -1\right)  r^{3}%
}=\kappa\left(  \rho\left(  r\right)  +p_{r}\left(  r\right)  +\tau_{r}\left(
r\right)  \right)  .
\end{equation}
At the throat, we find%
\begin{equation}
\label{nec_last}\frac{4-5r_{0}^{2}\left(  \Omega-\Omega_{0}\right)  ^{2}%
\sin^{2}\left(  \theta\right)  }{3r_{0}^{2}\left(  \left(  \Omega-\Omega
_{0}\right)  ^{2}r_{0}^{2}\sin^{2}\left(  \theta\right)  -1\right)  }%
=\kappa\left(  \rho\left(  r_{0}\right)  +p_{r}\left(  r_{0}\right)  \right)
.
\end{equation}
For the ZAMO frame, we get%
\begin{equation}
-\frac{4}{3r_{0}^{2}}=\kappa\left(  \rho\left(  r_{0}\right)  +p_{r}\left(
r_{0}\right)  \right)
\end{equation}
confirming the violation of the NEC. This time%
\begin{equation}
\lim_{r \rightarrow r_{0}}\lim_{\Omega\rightarrow\Omega_{0}}\kappa\left(
\rho\left(  r\right)  +p_{r}\left(  r\right)  \right)  =\lim_{\Omega
\rightarrow\Omega_{0}}\lim_{r \rightarrow r_{0}}\kappa\left(  \rho\left(
r\right)  +p_{r}\left(  r\right)  \right)  .\label{C}%
\end{equation}
For this TW as well, it can be showen that $\rho\left(  r_{0}\right)
+p_{r}\left(  r_{0}\right)  =0$ outside the ZAMO frame. Indeed, from
\eqref{nec_last}, we get
\begin{equation}
\frac{4-5r_{0}^{2}\left(  \Omega-\Omega_{0}\right)  ^{2}\sin^{2}\left(
\theta\right)  }{3r_{0}^{2}\left(  \left(  \Omega-\Omega_{0}\right)  ^{2}%
r_{0}^{2}\sin^{2}\left(  \theta\right)  -1\right)  }=0 \,.
\end{equation}
The numerator vanishes for
\begin{equation}
\Omega_{5,6}=\Omega_{0}\pm\frac{2\sqrt{5}}{5r_{0}\sin\left(  \theta\right)  }%
\end{equation}
and can be compared with \eqref{N1}. On the other hand, the denominator
vanishes for
\begin{equation}
\Omega_{7,8}=\Omega_{0}\pm\frac{1}{r_{0}\sin\left(  \theta\right)  } \,.
\end{equation}
This means that for the folowing values of $\Omega$
\begin{align}
\Omega_{7}  &  >\Omega>\Omega_{5}\\
\Omega_{6}  &  >\Omega>\Omega_{8} \,,
\end{align}
the NEC is not violated. For the fixed plates case, it is possible to extract
additional information on the angular velocity from the definition of the
ergosurface. Indeed, this occurs when
\begin{equation}
1=rK(r,{\theta})\Omega(r,{\theta})\sin{\theta=r\Omega\sin{\theta}}%
\end{equation}
where we have used \eqref{RFfp} and $\Omega\left(  r\right)  =\Omega\,$. To
stay outside the ergoregion, we must impose the following inequality%
\begin{equation}
1>r\Omega\sin{\theta} \,.
\end{equation}
At the throat this becomes
\begin{equation}
\frac{1}{r_{0}\sin{\theta}}>\Omega\quad\Longrightarrow\quad\frac{3r_{1}}%
{\sqrt{3}d^{2}\sin{\theta}}>\Omega\label{Ergoi}%
\end{equation}
implying that the rotation of the TW is very small.

\section{Conclusions}

\label{p4}

In this paper, we have extended the study of Casimir Wormholes by including
rotations. Given the complexity of deriving entirely new solutions, we adopted
the following strategy: the powering source is assumed to be a Casimir device
generating a SET represented by \eqref{Casimir}. To ensure consistency, we
imposed that when rotation ceases, the static solution is recovered.
Consequently, we had to distinguish between two scenarios:

\begin{description}
\item[a)] the ordinary Casimir Wormhole with radially varying plates,

\item[b)] the TW powered by the SET \eqref{Casimir}, where the plates are
parametrically fixed.
\end{description}

For the ordinary Casimir wormhole, we found that a solution is represented by
a constant angular velocity $\Omega$. The SET is given by \eqref{CSET}, with
the inclusion of the thermal tensor where only the transverse component
survives, ensuring the consistency of the EFE. Nevertheless, the constant
angular velocity $\Omega$ has an unpleasant feature: it extends throughout the
entire space, implying that dragging effects remain detectable even at
infinite distances. To address this issue, we modified the angular velocity by
including a damping factor of exponential form. This adjustment suppresses the
rotational effects, thereby confining them to the vicinity of the throat. To
avoid a change in the signature, we restricted the analysis to the region
outside the ergoregion. This assumption also enabled us to establish an upper
bound for the angular velocity. Interestingly, on the ergosurface, the TW
exhibits ultra-spinning behavior. This property is a consequence of having a
Planckian radius for the throat, as described by \eqref{r0}. Regarding the
NEC, it is found to be violated at the throat, even if we adopt the ZAMO
frame. However, a form of noncommutative behavior arises between the throat
region and the ZAMO frame, as expressed by \eqref{NC}. Concerning case
\textbf{b)}, it is found that the static TW profile predicts a large throat
radius, as given by \eqref{r0d}, which is consistent with the result found in
Ref. \cite{GABTW}. Additionally, a natural external boundary appears close to
the throat. This implies that, in the rotating case \textbf{b)}, the constant
$\Omega$ solution does not require any modification. Regarding the NEC in this
second case, we have observed a commutative behavior, as described by
\eqref{C}. On the other hand, in the region outside the ergoregion, the
inequality \eqref{Ergoi} leads to a very slow rotation, which contrasts with
the prediction made in case \textbf{a)}. Finally, concerning the double limit
described by \eqref{NC} and \eqref{C}, at this stage of the investigation, the
origin of such behavior remains unknown.

\appendix

\section{Einstein Field Equations}

\label{App}

$G_{tt}=\kappa T_{tt}$:
\begin{align}
&  \left(  \frac{r_{1}^{2}}{\kappa\,r^{4}}+\tau_{t}\left(  r\right)  \right)
\left(  -e^{2\Phi\left(  r\right)  }+r^{2}\Omega\left(  r\right)  ^{2}\sin
^{2}\left(  \theta\right)  \right) \nonumber\\
&  +\frac{\left(  \tau_{\rho}\left(  r\right)  +\tau_{t}\left(  r\right)
\right)  \left(  e^{-\Phi\left(  r\right)  }\right)  ^{2}\left(
-e^{2\Phi\left(  r\right)  }+r^{2}\sin^{2}\left(  \theta\right)  \left(
\Omega\left(  r\right)  ^{2}-\Omega\left(  r\right)  \Omega_{0}\right)
\right)  ^{2}}{1-r^{2}\sin^{2}\left(  \theta\right)  \left(  \Omega\left(
r\right)  -\Omega_{0}\right)  ^{2}\left(  e^{-\Phi\left(  r\right)  }\right)
^{2}}\nonumber\\
&  =\frac{\sin^{2}\left(  \theta\right)  r^{2}}{\kappa}\left(  1-\frac
{b\left(  r\right)  }{r}\right)  \Omega\left(  r\right)  \left(  \Phi
^{\prime\prime}\left(  r\right)  \Omega\left(  r\right)  -\Omega^{\prime
\prime}\left(  r\right)  +\left(  \Phi^{\prime}\left(  r\right)  \right)
^{2}\Omega\left(  r\right)  \right) \nonumber\\
&  -\frac{3\sin^{2}\left(  \theta\right)  \left(  r-b\left(  r\right)
\right)  \left(  r^{2}\Omega\left(  r\right)  ^{2}\left(  \sin^{2}\left(
\theta\right)  \right)  e^{-2\Phi\left(  r\right)  }+\frac{1}{3}\right)
\left(  \Omega^{\prime}\left(  r\right)  \right)  ^{2}r}{4}\nonumber\\
&  +\frac{2b^{\prime}\left(  r\right)  e^{2\Phi\left(  r\right)  }%
-r\Omega\left(  r\right)  ^{2}\sin^{2}\left(  \theta\right)  \left(
b^{\prime}\left(  r\right)  r-b\left(  r\right)  \right)  }{2r^{2}}-\frac
{\sin^{2}\left(  \theta\right)  \Omega\left(  r\right)  ^{2}\left(  b^{\prime
}\left(  r\right)  r+b\left(  r\right)  -2r\right)  \Phi^{\prime}\left(
r\right)  }{2}\nonumber\\
&  +\Omega^{\prime}\left(  r\right)  \sin^{2}\left(  \theta\right)
\Omega\left(  r\right)  \left(  r\left(  r-b\left(  r\right)  \right)  \left(
\Phi^{\prime}\left(  r\right)  \right)  +\frac{b^{\prime}\left(  r\right)
r}{2}-4r+\frac{7b\left(  r\right)  }{2}\right) \label{GTtt}%
\end{align}

$G_{{rr}}=\kappa T_{{rr}}$:
\begin{gather}
 \frac{r^2 }{32\pi} \left(1 - \frac{b(r)}{r}\right) \left( r^3 \sin^2\theta \ (\Omega '(r))^2 e^{-2\Phi(r)} +8\Phi '(r) \right) - \frac{b(r)}{8\pi} = \tau_{ r}(r) r^3- \frac{3r_1^2}{\kappa r}
 \nonumber\\
\label{GTrr}%
\end{gather}

$G_{{\theta\theta}}=\kappa T_{{\theta\theta}}$:
\begin{gather}
\left(  1-\frac{b\left(  r\right)  }{r}\right)  \Phi^{\prime\prime}\left(
r\right)  +\left(  1-\frac{b\left(  r\right)  }{r}\right)  \left(
\Phi^{\prime}\left(  r\right)  \right)  ^{2} +\frac{\Phi^{\prime}\left(
r\right)  }{2r}\left(  \left(  1-\frac{b\left(  r\right)  }{r}\right)
+\left(  1-b^{\prime}\left(  r\right)  \right)  \right) \nonumber\\
-\frac{r^{4}\sin^{2}\left(  \theta\right)  e^{-2\Phi\left(  r\right)  }%
}{4r^{2}}\left(  \Omega^{\prime}\left(  r\right)  \right)  ^{2}\left(
1-\frac{b\left(  r\right)  }{r}\right)  +\frac{b\left(  r\right)  -b^{\prime
}\left(  r\right)  r}{2r^{3}}=\left(  \frac{r_{1}^{2}}{r^{4}}+\kappa\tau
_{t}\left(  r\right)  \right) \label{GTthth}%
\end{gather}

$G_{\varphi\varphi}=\kappa T_{\varphi\varphi}$:%

\begin{gather}
\frac{r^{2}\sin^{2}\left(  \theta\right)  }{\kappa}\left(  1-\frac{b\left(
r\right)  }{r}\right)  \left(  \Phi^{\prime\prime}\left(  r\right)
-\frac{3\left(  \sin^{2}\left(  \theta\right)  \right)  e^{-2\Phi\left(
r\right)  }r^{2}\left(  \Omega^{\prime}\left(  r\right)  \right)  ^{2}}%
{4}+\left(  \Phi^{\prime}\left(  r\right)  \right)  ^{2}+\frac{1}{r}%
\Phi^{\prime}\left(  r\right)  \right) \nonumber\\
+\frac{\left(  r\left(  \Phi^{\prime}\left(  r\right)  \right)  +1\right)
}{2\kappa r}\left(  b\left(  r\right)  -\left(  b^{\prime}\left(  r\right)
\right)  r\right)  =\left(  \frac{r_{1}^{2}}{\kappa r^{4}}+\tau_{t}\left(
r\right)  \right)  r^{2}\left(  \sin^{2}\left(  \theta\right)  \right)
\nonumber\\
+\frac{\left(  \tau_{\rho}\left(  r\right)  +\tau_{t}\left(  r\right)
\right)  \left(  e^{-\Phi\left(  r\right)  }\right)  ^{2}\left(  \Omega\left(
r\right)  -\Omega_{0}\right)  ^{2}r^{4}\left(  \sin^{4}\left(  \theta\right)
\right)  }{1-r^{2}\left(  \sin^{2}\left(  \theta\right)  \right)  \left(
\Omega\left(  r\right)  -\Omega_{0}\right)  ^{2}\left(  e^{-\Phi\left(
r\right)  }\right)  ^{2}}\label{GTpp}%
\end{gather}

$G_{t\varphi}=\kappa T_{t\varphi}$:%

\begin{align}
&  -\left(  \frac{r_{1}^{2}}{\kappa\,r^{4}}+\tau_{t}\left(  r\right)  \right)
r^{2}\Omega\left(  r\right)  \sin^{2}\left(  \theta\right) \nonumber\\
&  +\frac{\left(  \tau_{\rho}\left(  r\right)  +\tau_{t}\left(  r\right)
\right)  \left(  e^{-\Phi\left(  r\right)  }\right)  ^{2}\left(
-e^{2\Phi\left(  r\right)  }+r^{2}\sin^{2}\left(  \theta\right)  \left(
\Omega\left(  r\right)  ^{2}-\Omega\left(  r\right)  \Omega_{0}\right)
\right)  \left(  \Omega\left(  r\right)  -\Omega_{0}\right)  r^{2}\sin
^{2}\left(  \theta\right)  }{1-r^{2}\sin^{2}\left(  \theta\right)  \left(
\Omega\left(  r\right)  -\Omega_{0}\right)  ^{2}\left(  e^{-\Phi\left(
r\right)  }\right)  ^{2}}\nonumber\\
&  =\frac{r^{2}\sin^{2}\left(  \theta\right)  }{\kappa}\left(  1-\frac
{b\left(  r\right)  }{r}\right)  \left(  -\Omega\left(  r\right)  \Phi
^{\prime\prime}\left(  r\right)  +\frac{\Omega^{\prime\prime}\left(  r\right)
}{2}-\Omega\left(  r\right)  \left(  \Phi^{\prime}\left(  r\right)  \right)
^{2}\right)  -\frac{3\sin^{2}\left(  \theta\right)  e^{-2\Phi\left(  r\right)
}\Omega\left(  r\right)  r^{2}\left(  \Omega^{\prime}\left(  r\right)
\right)  ^{2}}{4}\nonumber\\
&  -\frac{\sin^{2}\left(  \theta\right)  \Omega^{\prime}\left(  r\right)
}{4\kappa}\left(  2r^{2}\left(  1-\frac{b\left(  r\right)  }{r}\right)
\Phi^{\prime}\left(  r\right)  +\left(  b^{\prime}\left(  r\right)  \right)
r-8r+7b\left(  r\right)  \right) \nonumber\\
&  -\frac{\sin^{2}\left(  \theta\right)  \Omega\left(  r\right)  }{2\kappa
r}\left(  \left(  r\left(  \left(  b^{\prime}\left(  r\right)  \right)
r-2r+b\left(  r\right)  \right)  \Phi^{\prime}\left(  r\right)  +\left(
b^{\prime}\left(  r\right)  \right)  r-b\left(  r\right)  \right)  \right)
\label{GTtp}%
\end{align}

\end{document}